\documentclass [12 pt, a4paper]{article}
\usepackage[cp1251]{inputenc}
\usepackage[T2A]{fontenc}
\usepackage{amssymb,amsmath}
\usepackage[english]{babel}
\RequirePackage[dvips]{graphicx}

\begin{document}
\begin{center}
\textbf{SYNCHRONIZATION OF COUPLED ANIZOCHRONOUS OSCILLATORS}\\
\medskip{Alexander~P.~Kuznetsov$^{a}$, Julia~P.~Roman$^{b}$}\\
\small{\it $^{a}$Institute of Radio-Engineering and Electronics of RAS, Zelenaya 38, Saratov 410019, Russia\\
$^{b}$Department of Nonlinear Processes, Saratov State University, Astrakhanskaya 83, Saratov 410026, Russia}
\end{center}

\begin{abstract}
The particular properties of synchronization are discussed for the system of coupled van der Pol -- Duffing oscillators. The arrangement of synchronization tongues and the particular properties of their internal structure in the parameter space are revealed. The features of attractors in the phase space and in the Poincare section are consi\-dered.
\end{abstract}

\renewcommand{\thesection}{\normalsize{\arabic{section}.}}
\section{\normalsize{Introduction}}

\hspace*{\parindent}The system of two coupled van der Pol oscillators is one of the basic mo\-dels of nonlinear dynamics demonstrating the phenomenon of mutual synchronization. There are many papers on this theme because this system demonstrates a lot of interesting oscillation regimes and types of behavior, such as synchronous and quasiperiodic regimes, synchronization with different phase conditions [Pikovsky et al., 2001], the "oscillation death" effect [Aronson et al., 1990; Pikovsky et al., 2001], global bifurcations [Chakraborty \& Rand, 1988], etc. Moreover, the results of this system investigation may be used in the analysis of different electronic, biological and chemical systems [Cohen \& Neu, 1979; Minorsky, 1962; Neu, 1979; Pavlidis, 1973; Poliashenko et al., 1991]. Interest in this problem does not decrease because of its signi\-ficance and dynamics variety. In this respect we may point out, for example, the paper [Ivanchenko et al., 2004], where the resumptive study of this system is undertaken within the framework of quasiharmonic approximation. In particular, they took into account the possible non-identity of oscillators and combined nature of coupling.

The system of differential equations describing the interaction between van der Pol oscillators is of the form
\begin{equation}
\begin{split}
&\frac{\displaystyle d^{2}x}{\displaystyle dt^{2}}-(\lambda-x^{2})\frac{\displaystyle dx}{\displaystyle dt}+x+\mu(\frac{\displaystyle dx}{\displaystyle dt}-\frac{\displaystyle dy}{\displaystyle\displaystyle dt})=0,\\
&\frac{\displaystyle d^{2}y}{\displaystyle dt^{2}}-(\lambda-y^{2})\frac{\displaystyle dy}{\displaystyle dt}+(1+\delta)y+\mu(\frac{\displaystyle dy}{\displaystyle dt}-\frac{\displaystyle dx}{\displaystyle dt})=0.\\
\end{split}
\end{equation}

Here $\lambda$ is parameter characterizing the excess above the threshold of the Andronov -- Hopf bifurcation in autonomous oscillators, $\delta$ is the frequency mismatch between the second and first oscillators, and $\mu$ is the coefficient of dissipative coupling.

Investigation of the system of interacting autooscillatory systems may be carried out on different "levels of sophistication": by means of Adler equation for oscillators phase difference dynamics [Pikovsky et al., 2001; Rand \& Holmes, 1980], by means of abridged equations, which are correct in quasiharmonic approximation, when oscillators exceed slightly the threshold of the Andronov -- Hopf bifurcation [Aronson et al., 1990; Ivanchenko et al., 2004; Pikovsky et al., 2001], and, at last, by means of initial Eqs. (1) [Chakraborty \& Rand ,1988; Pastor-Diaz \& Lopez-Fraguas, 1995; Storti \& Rand, 1982]. The last is the most difficult task due to both more complicated behavior of the initial system (1) and the solution dependence on the additional para\-meter $\lambda$\footnote{For abridged equations this parameter is not essential and can be eliminated due to renormalization [Pikovsky et al., 2001].}.

When studying the dynamics of excited and coupled systems it is important to take into account anisochronism of small oscillations. If we talk about the phase dynamics of individual oscillator, anisochronism is the dependence of the rate of phase change on oscillation amplitude. This effect may be considered, if we insert additional cubic nonlinearity of Duffing oscillator type into Eq. (1). Then the equations for coupled oscillators are
\begin{equation}
\begin{split}
&\frac{\displaystyle d^{2}x}{\displaystyle dt^{2}}-(\lambda-x^{2})\frac{\displaystyle dx}{\displaystyle dt}+x+\beta x^{3}+\mu(\frac{\displaystyle dx}{\displaystyle dt}-\frac{\displaystyle dy}{\displaystyle\displaystyle dt})=0,\\
&\frac{\displaystyle d^{2}y}{\displaystyle dt^{2}}-(\lambda-y^{2})\frac{\displaystyle dy}{\displaystyle dt}+(1+\delta)y+\beta y^{3}+\mu(\frac{\displaystyle dy}{\displaystyle dt}-\frac{\displaystyle dx}{\displaystyle dt})=0.\\
\end{split}
\end{equation}
The parameter $\beta$ is responsible for anisochronism of small oscillations\footnote{In system of Eqs. (1) anisochronism is possible also due to not small values of parameter $\lambda$. But we will call for convenience parameter $\beta$ also the parameter of phase nonlinearity or anisochronism.}. To demonstrate this in an explicit form we give the corresponding system of abridged equations for the amplitudes of oscillators $R$ and $r$ and the difference of their phases $\varphi$ [Pikovsky et al., 2001], which may be derived from (2) in the quasiharmonic approximation:
\begin{equation}
\begin{split}
&\frac{\displaystyle dR}{\displaystyle dt}=R(\lambda-\mu)-R^{3}+\mu r \cos \varphi,\\
&\frac{\displaystyle dr}{\displaystyle dt}=r(\lambda-\mu)-r^{3}+\mu R \cos \varphi,\\
&\frac{\displaystyle d\varphi}{\displaystyle dt}=\delta+3\beta (r^{2}-R^{2})-\mu(\frac{\displaystyle r}{\displaystyle R}+\frac{\displaystyle R}{\displaystyle r})\sin \varphi.\\
\end{split}
\end{equation}
One can see that the rate of change of oscillators' phase difference $\varphi$ depends on the difference of squared amplitudes $R$ and $r$ even if the coupling is absent.

There are two reasons why taking anisochronism into account is important. First, with taking this factor into account the system equations may be brought to the normal form of Andronov -- Hopf bifurcation in individual oscillator [Pikovsky et al., 2001]. In these regards the van der Pol system is partly "degenerated" and only the transition to the van der Pol -- Duf\-fing system removes this degeneracy. At the same time choice of the system which can be reduced to the normal form of Andronov -- Hopf bifurcation is important with regard to generalizations and to application of the results to another systems.

Second, it is known that for the case of synchronization by external force the anisochronism leads to new dynamics features. It is more obvious for the systems with pulse excitation. In that case the isochronous system leads to the map for the phase, which was studied by Ding [1986-1988], Glass et. al. [1983], Glass \& Sun [1994], Keener \& Glass [1984], Ullmann \& Caldas [1996], Viana \& Batista [1998]:
\begin{equation}
\varphi_{n+1}=\arctan\left(\frac{\sin\varphi_{n}+C}{\cos\varphi_{n}}\right)+T.
\end{equation}
Here $C$ and $T$ are amplitude and period of influence correspondingly. In the case of strong anisochronism it may be reduced to the standard circle map [Pikovsky et al., 2001]:
\begin{equation}
\theta_{n+1}=\theta_{n}+\Omega-3\beta C \sin\theta_{n},
\end{equation}
where $\Omega=T(1+3\beta\lambda/2)$ is normalized period accounting for additional phase lag.

These two types of maps give two pictures of synchronization tongues different in many respects especially in internal structure of these tongues [Ding, 1986-1988; Glass et. al., 1983; Glass \& Sun, 1994; Keener \& Glass, 1984; Pikovsky et al., 2001; Ullmann \& Caldas, 1996; Viana \& Batista, 1998]. This signalizes that the arrangement of synchronization tongues and their internal structure will demonstrate essential peculiarities for coupled anisochronous systems too. At the same time the system of coupled van der Pol -- Duffing oscillators (2) practically was not investigated in this context. One should emphasize that quasiharmonic approximation [Aronson et al., 1990; Ivanchenko et al., 2004; Pikovsky et al., 2001] is not efficient in that case, since it does not describe higher order synchronization tongues and the possibility of period doublings and chaos inside them. In the present paper we shall give some results of investigation of system (2).

\section{\normalsize{Structure of the parameters space and illustrations of dynamics}}

\hspace*{\parindent}Sufficiently simple and informative method of studying the parameters space structure of nonlinear systems is the method of dynamic regime chart construction [Kuznetsov et al., 1997; Kuznetsov et al., 2006]. Within the framework of such a method we shall mark the oscillation period of the system of coupled oscillators by means of different colors on the parameter plane (frequency mismatch $\delta$ -- coupling value $\mu$). White color corresponds to the chaotic or quasiperiodic motions. Cycles periods were calculated by means of the Poincare section method: this is the number of points of intersection of the phase trajectory on the attractor and the surface chosen as the Poincare section. Only those crossings were taken into account that correspond to the trajectories coming to the surface from the one side.

The system under investigation is characterized by four-dimensional phase space $(x,\dot{x},y,\dot{y})$. Therefore three-dimensional hypersurface that is preset by means of some additional condition, e.g., zero velocity of the second oscillator $\dot{y}=0$ may serve as the Poincare section. In that case the number $n$ of points of intersection of trajectory and the hypersurface was determined. Colors on the charts are chosen in accordance with the period $n$.

The chart of dynamic regimes obtained in such a way for the van der Pol oscillators (1) is given in Fig.\:1(a) on the $(\delta, \mu)$-plane for $\lambda=2.5$. The value of the control parameter is chosen large, that quasiharmonic approximation is not efficient.

We shall characterize mutual oscillations of oscillators also by means of the rotation number $w$. By analogy with [Kuznetsov et al., 2006; Postnov et al., 2005] we find numerically the average return time $\tau_{y}$  for the chosen Poincare section $\dot{y}=0$ and the average return time $\tau_{x}$ for the section of the first oscillator defined by $x=0$. Then the rotation number is defined as $w=\tau_{y}/\tau_{x}$. On the chart in Fig.\:1(a) there are shown: main synchronization tongue with the frequency ratio of 1/1; area of quasiperiodic regimes with the embedded system of higher order synchronization tongues among which the tongues with the rotation numbers of 2/3 and 3/5 are the most characteristic; the area of the "oscillation death" effect [Aronson et al., 1990; Pikovsky et al., 2001] which corresponds to stabilization of the equilibrium state point at the origin due to sufficiently strong dissipative coupling.

\begin{figure}[h!]
\begin{center}
\includegraphics[scale=0.9]{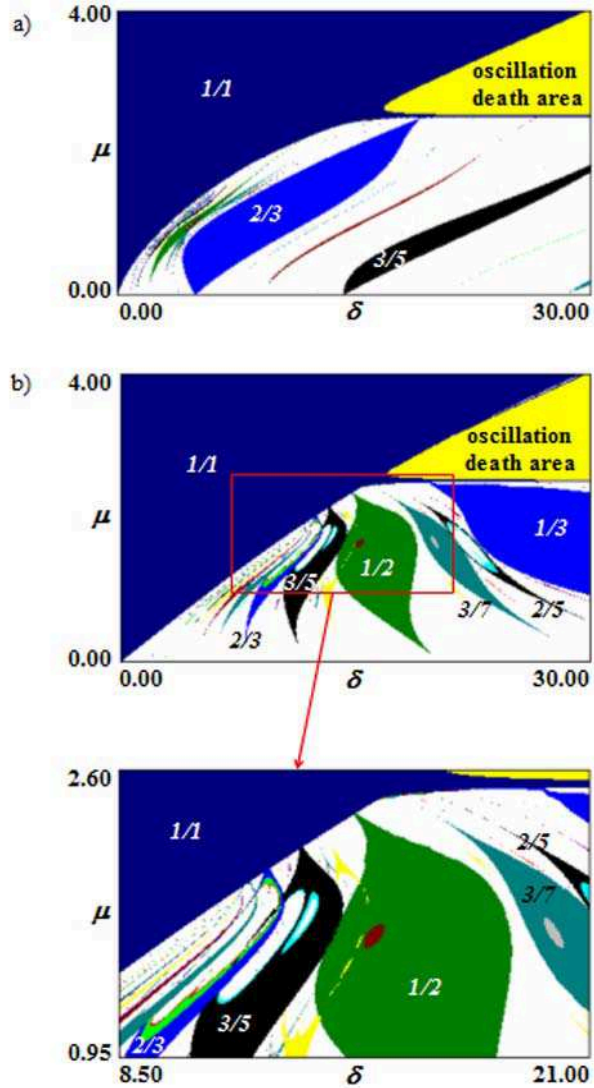}
\end{center}
\raggedright
\caption{Dynamic regime charts for the system (2) for $\lambda=2.5$ in case of (a)~isochronous $(\beta=0)$, and (b, c) anisochronous $(\beta=1)$ oscillators.}
\end{figure}

Now we pass on to the van der Pol -- Duffing model (2). Corresponding chart of dynamic regimes and its fragment are shown in Fig.\:1(b, c) for $\beta=1$ that meet the case of sufficiently great anisochronism.

The graph of the rotation number $w$ versus the frequency mismatch $\delta$ is shown for this case in Fig.\:2(a) for different values of the coupling parameter. On this graph one can see characteristic wide "steps" that correspond to different rational rotation numbers. When coupling parameter is not great $(\mu=0.5)$ steps are feebly marked. This means that quasiperiodic regimes predominate. With growth of the coupling parameter $(\mu=1)$ a number of sufficiently wide steps emerge. They correspond (in decreasing order) to the rotation number values $w$ = 1/2, 3/5, 2/3, etc. If we continue to increase the value of the coupling parameter $(\mu=2.3)$, synchronization regime with the rotation number of 1/3 begins to predominate.

\begin{figure}[h!]
\begin{center}
\includegraphics[scale=0.6]{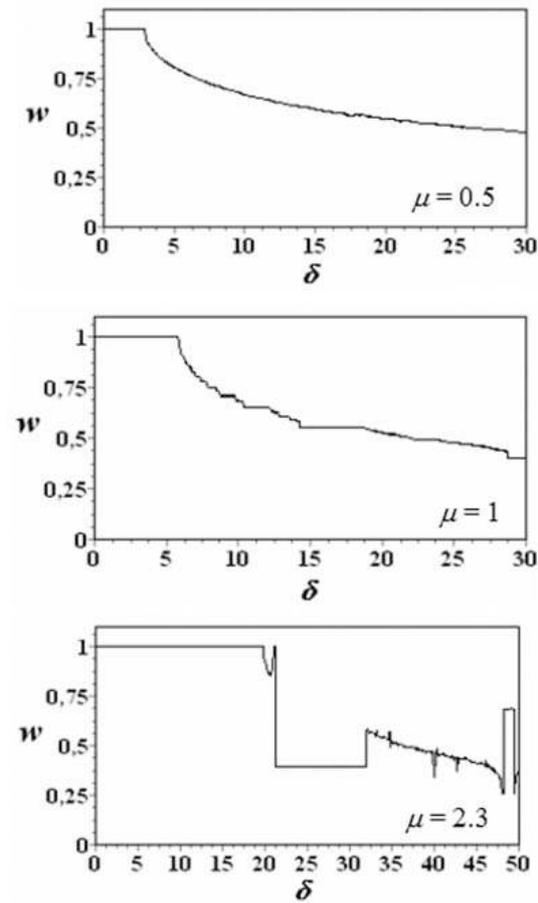}
\end{center}
\raggedright
\caption{Dependences of the rotation number $w$ upon the frequency mismatch $\delta$ for $\lambda=2.5$, $\beta=1$ and different values of the coupling parameter $\mu$.}
\end{figure}

One can see in Fig.\:1(b) that the presence of anisochronism leads to the displacement of synchronization tongues towards the greater values of frequency mismatch. Synchronization tongues become so wide that one can see the situation of their overlapping, which is characteristic for standard circle map. Internal structure of this tongues changes too: period doublings and transition to chaotic dynamics inside them are observed. It may be seen clearly in enlarged fragment of the dynamic regime chart, shown in Fig.\:1(c).

With the increase of the parameter of anisochronism $\beta$ an onset of new "islands" of doubled periods takes place inside the synchronization tongues locating at greater values of frequency mismatch $\delta$ (see Fig.\:3).
\begin{figure}[h!]
\begin{center}
\includegraphics[scale=0.5]{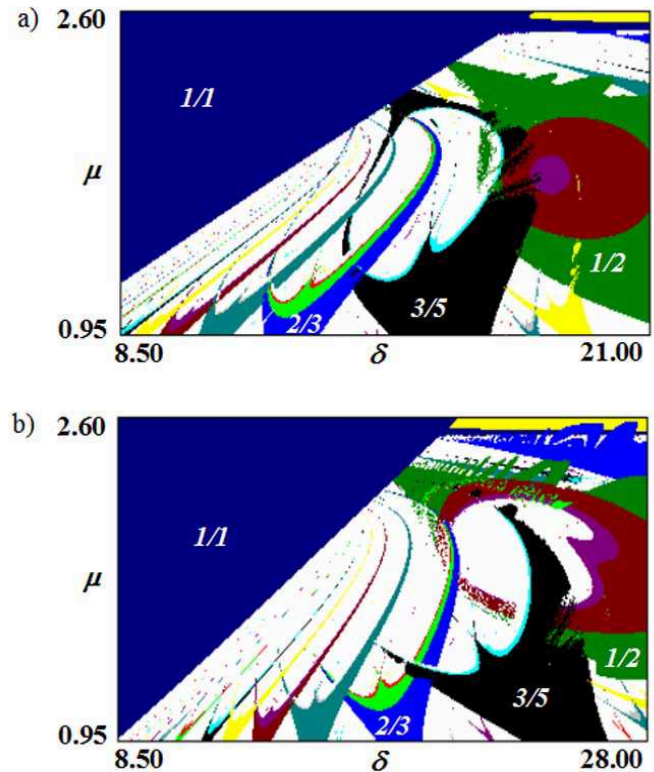}
\end{center}
\raggedright
\caption{Dynamic regime charts for the system (2) for $\lambda=2.5$ and (a) $\beta=1.7$; (b) $\beta=3$.}
\end{figure}
Synchronization tongues change their shape with the further increase of parameter $\beta$. They fall into two similar in structure pieces, inside which structures called "crossroad area" exist [Carcasses et al., 1991; Kuznetsov et al., 1997]. These structures are characteristic for systems with period doublings. Practically, two systems of synchronization tongues may be observed in the parameters plane: tops of the first one are disposed along the line of zero coupling, tops of the second one -- along the boundary of main synchronization area 1/1, which is the line of Neimark -- Sacker bifurcation. Chaotic area appears between these two systems of synchronization tongues.

When we analyze chaotic areas it is useful to supplement dynamic regime charts with the chart of the first Lyapunov exponent (see Fig.\:4). Blue color shadings correspond on this chart to the synchronization areas with negative Lyapunov exponent. Deeper color corresponds to greater absolute value of the exponent. Yellow and orange colors hues correspond to exponent values close to zero and designate the areas of quasiperiodic regimes. Due to finite accuracy of numerical simulations it is not possible to detect perfectly zero value of the first Lyapunov exponent. Therefore areas of quasiperiodic regimes have slightly varying shadings. Grey color marks the oscillation death area. Red color corresponds to chaotic regimes with positive Lyapunov exponent. The existence of red color on the chart of the first Lyapunov exponent verifies presence of chaotic behavior in the system under investigation.

\begin{figure}[h!]
\begin{center}
\includegraphics[scale=0.5]{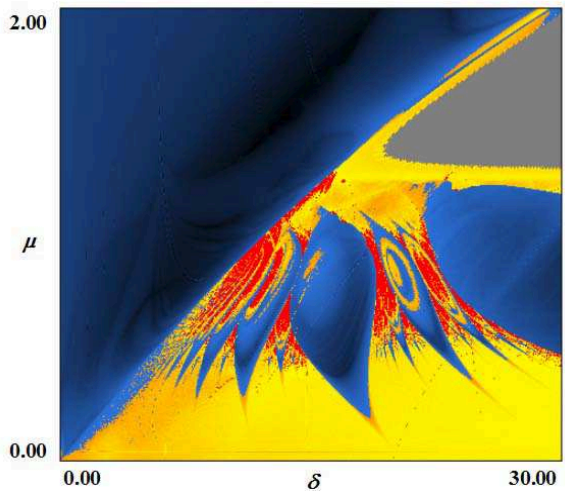}
\end{center}
\raggedright
\caption{Chart of the first Lyapunov coefficient for the system of coupled van der Pol -- Duffing oscillators for $\lambda=2.5$, $\beta=1$.}
\end{figure}

Phase portraits of attractors computed at several characteristic points within domains of quasiperiodic regimes are given in Fig.\:5 on $(x,\dot{x})$ and $(y,\dot{y})$ planes for the first and the second oscillators correspondingly. To put it more precisely these portraits are projections of the attractor in phase space $(x,\dot{x},y,\dot{y})$ onto corresponding planes $(x,\dot{x})$ and $(y,\dot{y})$. 
\begin{figure}[h!]
\begin{center}
\includegraphics[scale=1]{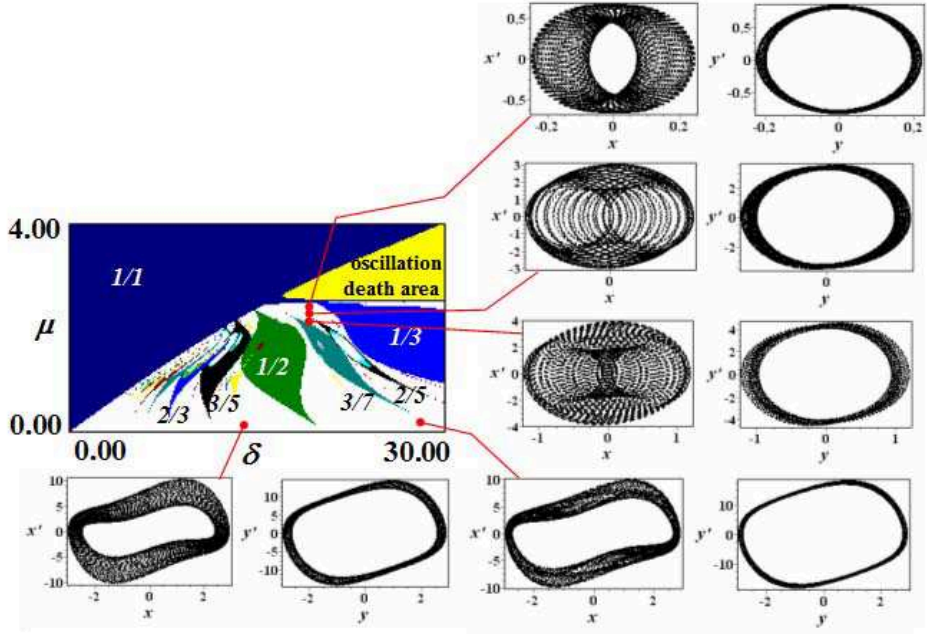}
\end{center}
\raggedright
\caption{Phase plane portraits at characteristic points of area of quasiperiodic regimes on the dynamic regime chart for $\lambda=2.5$, $\beta=1$.}
\end{figure}
For small values of coupling parameter it may be seen that portraits of attractors look like slightly disturbed limit cycles of individual oscillators. At the same time the trajectory never approaches the neighborhood of the point of origin. With the increase of the coupling parameter the trajectories of individual oscillators become more perturbed and there exist a kind of a threshold value when the trajectory of the first oscillator may achieve the neighborhood of the origin. Corresponding phase portrait looks like the area completely "filled" with trajectories. Evidently, this situation corresponds to the case of ill-determined phase of this oscillator. When approaching the oscillation death area an essential downsizing of attractors occurs (see scales in coordinate axis). The orbit ceases again to achieve neighborhood of the origin, but structure of phase portraits and shapes of attractors are different from the case of small coupling value.

Three-dimensional illustrations of the dynamics in the "Poincare sections" are given in Fig.\:6. They illustrate dynamics of the system of coupled oscillators at the points in the parameters space inside the domains of quasiperiodic regimes and near the threshold of chaos. We should remind that the Poincare section was chosen as the hypersurface $\dot{y}=0$ in four-dimensional phase space $(x,\dot{x},y,\dot{y})$. Hence the dynamics of the Poincare return map may be represented by the orbit in three-dimensional space $(x,\dot{x},y)$ as is shown in Fig.\:6. One can see that phase portraits in such Poincare section are invariant curves, which illustrates quasiperiodic character of dynamics. When approaching the chaotic area the curves undergo deformations and break down via loss of smoothness.

\begin{figure}[h!]
\begin{center}
\includegraphics[scale=1]{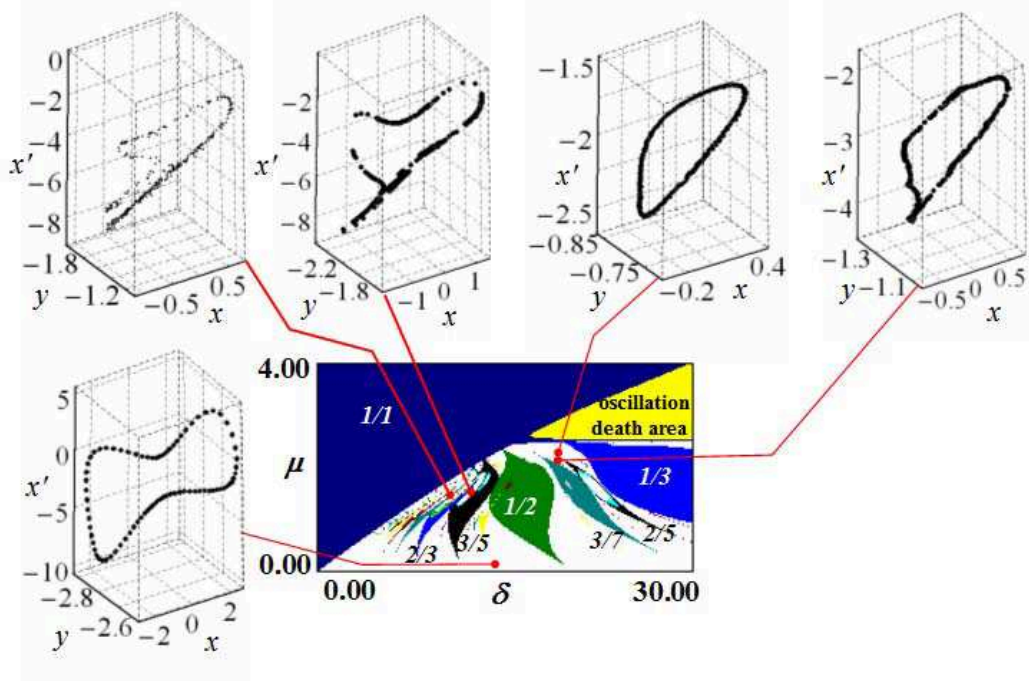}
\end{center}
\raggedright
\caption{Three-dimensional "Poincare sections" for the system (2) for \mbox{$\lambda=2.5$}, $\beta=1$.}
\end{figure}

Fig.\:7 illustrates the dependence of synchronization picture on the parameter of anisochronism $\beta$. There are presented dynamic regime charts on the parameter plane (frequency mismatch $\delta$ -- parameter of anisochronism $\beta$) at fixed coupling parameter values. One can see the "fan pattern" of high order synchronization tongues spreading away from the main synchronization tongue. With the increase of the coupling parameter these tongues widen and onset of "islands" of doubled periods takes place inside them. With further increase of coupling parameter "crossroad area" structures emerge inside synchronization tongues. Phase portraits computed at characteristic points of the dynamic regime chart are given in Fig.\:7(c). These portraits illustrate cascade of period doubling and appearance of chaos.

\begin{figure}[h!]
\begin{center}
\includegraphics[scale=1.5]{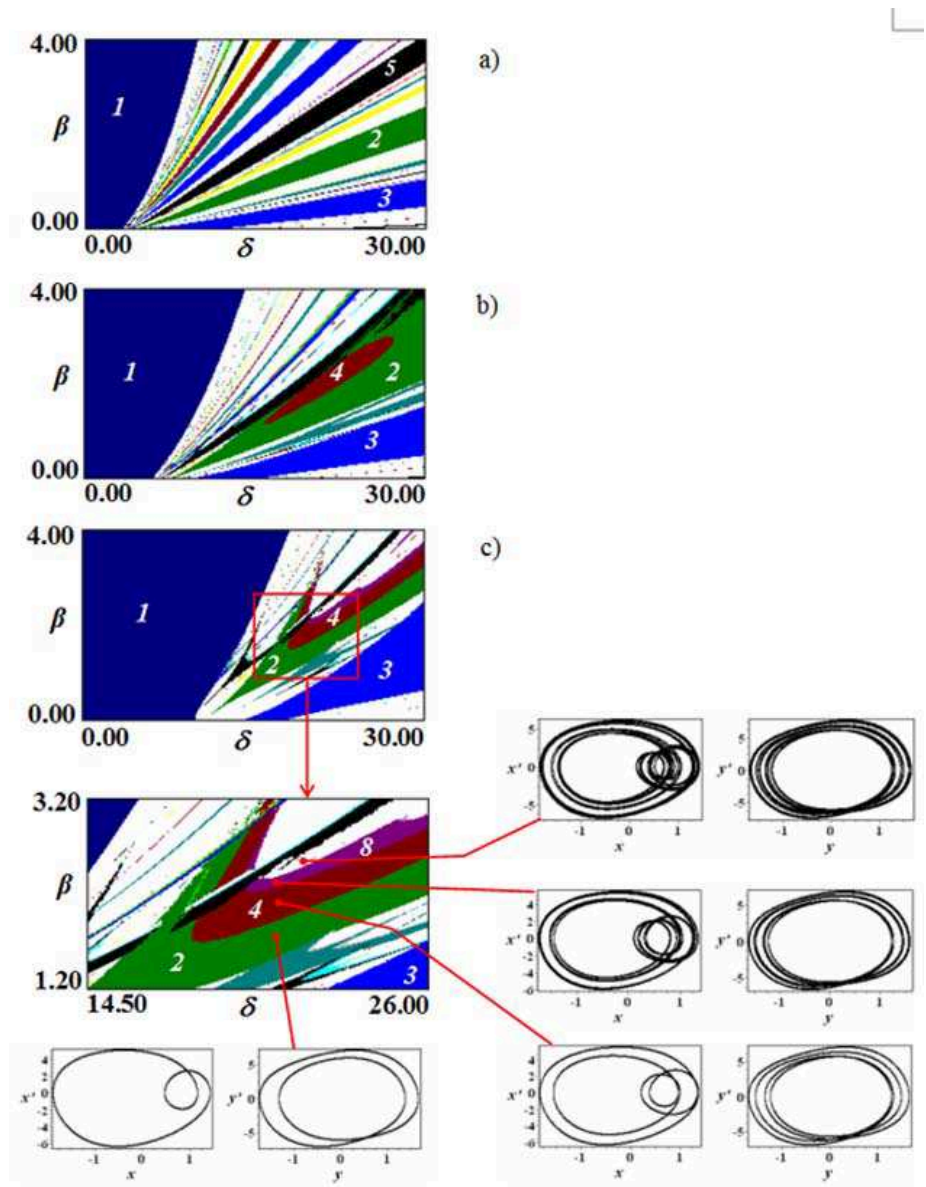}
\end{center}
\raggedright
\caption{Dynamic regime charts for the system (2) for $\lambda=2.5$ and (a) $\mu=1$; (b) $\mu=1.5$; (c) $\mu=2$, and phase plane portraits computed at typical areas.}
\end{figure}

\section{\normalsize{Conclusions}}

\hspace*{\parindent}The analysis of synchronization in the system of coupled anisochronous van der Pol -- Duffing oscillators reveals an essential role that the anisochronism plays in the formation of the picture of emerging regimes. Namely, in the present paper we have shown that synchronization tongues become essentially wider and the situation of their overlapping occurs. Islands of doubled periods, "crossroad area" structures and chaotic areas appear inside the synchronization tongues. In three-dimensional Poincare section we have observed invariant curves which may break down via the loss of smoothness with the formation of chaotic attractors.

\section*{\normalsize{Acknowledgements}}
\hspace*{\parindent}This work was supported by RFBR (Project No. 06-02-16773) and foundation of nonprofit programs "Dynasty".

\section*{\normalsize{References}}

\hspace*{\parindent}Aronson,~D.~G., Ermentrout,~G.~B., Kopell,~N. [1990] "Amplitude response of coupled oscillators", {\it Physica} {\bf D41}, 403-449.

Carcasses, J., Mira, C., Bosch, M., Simo, C. \& Tatjer, J.~C. [1991] " Crossroad area - spring area transition (I) parameter plane representation", {\it Int. J. Bifurcation and Chaos} {\bf 1}(1), 183-196.

Chakraborty, T. [1986] "Bifurcation analysis of two weakly coupled van der Pol oscillators", {\it Doctoral thesis} (Cornell University).

Chakraborty, T., Rand, R.~H. [1988] "The transition from phase locking to drift in a system of two weakly coupled van der Pol oscillators", {\it Int. J. Non-Linear Mechanics} {\bf 23}(5/6), 369-376.

Cohen, D.~S., Neu, J.~C. [1979] "Interacting oscillatory chemical reactors", in {\it Bifurcation Theory and Applications in the Scientific Disciplines}, eds. Gurel, O. \& Rossler, O. E. (Ann. N.Y. Acad. Sci. 316) pp. 332-337.

Ding, E.~J. [1986] "Analytic treatment of periodic orbit systematics for a nonlinear driven oscillator", {\it Phys. Rev.} {\bf A34}(4), 3547-3550.

Ding, E.~J. [1987] "Analytic treatment of a driven oscillator with a limit cycle", {\it Phys. Rev.} {\bf A35}(6), 2669-2683.

Ding, E.~J. [1987] "Structure of parameter space for a prototype nonlinear oscillator", {\it Phys. Rev.} {\bf A36}(3), 1488-1491.

Ding, E.~J. [1988] "Structure of the parameter space for the van der Pol oscillator", {\it Physica Scripta} {\bf 38}, 9-16.

Glass, L. et. al. [1983] "Global bifurcations of a periodically forced biological oscillator", {\it Phys. Rev.} {\bf A29}, 1348-1357.

Glass, L., Sun, J. [1994] "Periodic forcing of a limit-cycle oscillator: fixed points, Arnold tongues, and the global organization of bifurcations", {\it Phys. Rev.} {\bf 50}(6), 5077-5084.

Ivanchenko, M.~V., Osipov, G.~V., Shalfeev, V.~D., Kurths, J. [2004] "Synchronization of two non-scalar-coupled limit-cycle oscillators", {\it Physica} {\bf D189}, 8-30.

Keener, J.~P., Glass, L. [1984] "Global bifurcation of a periodically forced nonlinear oscillator", {\it J. Math. Biology} {\bf 21}, 175-190.

Kuznetsov, A.~P., Kuznetsov, S.~P., Sataev, I.~R. [1997] "A variety of period-doubling universality classes in multi-parameter analysis of transition to chaos", {\it Physica} {\bf D109}, 91-112.

Kuznetsov, A.~P., Mosekilde, E. \& Turukina, L.~V. [2006] "Synchronization in systems with bimodal dynamics", {\it Physica} {\bf A121}(2), 280-292.

Minorsky, N. [1962] {\it Nonlinear oscillators} (Van Nostrand).

Neu, J.~C. [1979] "Coupled chemical oscillators", {\it SIAM J. Appl. Math.} {\bf 37}(2), 307-315.

Pastor-Diaz, I., Lopez-Fraguas, A. [1995] "Dynamics of two coupled van der Pol oscillators", {\it Phys. Rev.} {\bf E52}, p. 1480.

Pavlidis, T. [1973] "Biological oscillators: the mathematical analysis", {\it Academic Press}.

Pikovsky, A., Rosenblum, M., Kurths, J. [2001] {\it Synchronization} (Cambridge), p. 411.

Poliashenko, M., McKay, S.~R., Smith, C.~W. [1991] "Hysteresis of synchronous - asynchronous regimes in a system of two coupled oscillators", {\it Phys. Rev.} {\bf A43}, p. 5638.

Poliashenko, M., McKay, S.~R., Smith, C.~W. [1991] "Chaos and nonisochronism in weakly coupled nonlinear oscillators", {\it Phys. Rev.} {\bf A44}, p. 3452.

Postnov, D.~E., Shishkin, A.~V., Sosnovtseva, O.~V. \& Mosekilde, E. [2005] "Two-mode chaos and its synchronization properties", {\it Phys. Rev.} {\bf E72}, pp. 056208(5).

Rand, R.~H., Holmes, P.~J. [1980] "Bifurcation of periodic motions in two weakly coupled van der Pol oscillators", {\it Int. J. Non-Linear Mechanics} {\bf 15}, 387-399.

Storti, D.~W., Rand, R.~H. [1982] "Dynamics of two strongly coupled van der Pol oscillators", {\it Int. J. Non-Linear Mechanics} {\bf 17}(3), 143-152.

Ullmann, K. \& Caldas, I.~L. [1996] "Transitions in the Parameter Space of a Periodically Forced Dissipative System", {\it Chaos, Solitons \& Fractals} {\bf 11}, p. 1913.

Viana, R.~L. \& Batista, A.~M. [1998] "Synchronization of coupled kicked limit cycle systems", {\it Chaos, Solitons \& Fractals} {\bf 9}(12), 1931-1944.

\end{document}